\documentclass[11pt,twoside]{article}

\usepackage{asp2006}
\usepackage{epsf}
\usepackage{psfig}
\usepackage{lscape}

\begin{document}
\title{Models for Dust and Molecular  Emission of High-Mass Protostars }   %%% Fill in title
\author{Mayra Osorio}   %%% Fill in author names
\affil{Instituto de Astrof\'\i sica de Andaluc\'\i a, CSIC,
Camino Bajo de Hu\'etor 50, E-18008 Granada, Spain; osorio@iaa.es}    %%% Fill in author affiliations

\begin{abstract} %%% Abstract to run on from here.

We present the results of a detailed modeling aimed to reproduce the 
spectral energy distribution (SED) of dust and molecular line emission of 
massive protostars under the hypothesis that they form via an accretion 
process. We model the emission originated in the infalling envelopes at 
scales smaller than 0.1 pc from the central protostar. To do that, we 
compared our model results with observational data covering a wide range 
of wavelengths, paying special attention to the high angular resolution 
mid-infrared data obtained with the Gemini Observatory and the ammonia 
line emission observed with the VLA at centimeter wavelengths. We have 
explored two kind of model envelopes.  In the first kind of models, 
spherical symmetry is assumed and the SED as well as the ammonia emission 
of the infalling envelope are calculated. In this way, the temperature, 
density, velocity, velocity dispersion, and ammonia abundance variations 
along the core can be obtained. The second approach takes into account 
deviations from the spherical symmetry, and parameters such as the 
rotation, degree of elongation of the core, or inclination of the system 
can be constrained through the SED fitting. Using these two approaches we 
have been able to model the formation of massive stars with a degree of 
detail similar to that reached for the low mass stars.

\end{abstract}

%%% MAIN BODY OF TEXT GOES HERE. CONSULT "INSTRUCTIONS FOR AUTHORS USING
%%% LATEX2E MARKUP", SECTIONS 2.3-2.6 FOR HELP WITH EQUATIONS, FIGURES,
%%% AND TABLES.

\section{Introduction}   
%\subsection{}   %%% Second level section head (remove "%" symbol)
%\subsubsection{}   %%% Lowest level section head (remove "%" symbol)
%\section*{}    %%% Unnumbered top level section head (remove "%" symbol)
%\subsection*{}   %%% Unnumbered second level section head (remove "%" symbol)

In this paper we summarize the results of some of our attempts to model 
the spectral energy distribution (SED) of dust as well as the molecular 
line emission of hot molecular cores (HMCs). Hot molecular cores are dense 
(10$^{6}-10^8$ cm$^{-3}$), hot (100-300 K), and small ($<5''$) 
condensations located in the vicinity of ultracompact HII (UCHII) regions. 
They are characterized by strong continuum millimeter dust emission, 
molecular lines of high excitation levels, and are frequently associated 
with water masers. Despite these physical conditions indicative of a 
powerful central source, HMCs are not associated with significant 
centimeter free-free emission. The study of this kind of sources has 
gained interest in the last years because it has been shown that they may 
be tracing a phase prior to the formation of an UCHII region, and 
therefore they may represent one of the earliest observable phases in the 
formation of a massive star.

The main hypothesis of our models is that massive stars are formed via an 
accretion process. Under this hypothesis a HMC is simulated as an envelope 
infalling onto a massive protostar, where the main source of energy is 
coming from the luminosity of the star and the accretion shock. For the 
geometry of the envelope two approaches are adopted. In a first 
approximation, spherical envelopes are assumed (Osorio, Lizano \& 
D'Alessio 1999) with a density distribution resulting from the collapse of 
the singular logatropic sphere (SLS, McLaughlin \& Pudritz 1997). For a 
given stellar mass and mass accretion rate the SLS collapse solution is 
able to determine the complete physical structure of the envelope. In the 
second approach, non-spherical envelopes are adopted. These envelopes are 
elongated not only in the inner region due to rotation but also at large 
radii due to intrinsic flattening of the core (De Buizer, Osorio \& Calvet 
2005). In addition to the dust emission, the molecular line emission is 
obtained for the physical parameters of the collapsing SLS derived from 
the fit to the observed SED (Osorio et al. 2007). The only free parameter 
in the line fitting is the gas-phase molecular abundance.

It is important to emphasize that to properly model the properties of HMCs 
high angular resolution observational data are required in order to 
distinguish the emission coming from the HMC from that of nearby UCHIIs. 
For this reason, in our modeling we use preferentially high angular 
resolution data. In the following sections we summarize briefly these 
models.

\section{Spherical Envelopes}  

Osorio et al. (1999) modeled the SED of several prototypical HMCs using 
mostly submillimeter and millimeter high angular resolution ($<5''$) data. 
For these sources only upper limits of the flux density were available at 
near- and mid-infrared wavelengths, so spatial intensity profiles at 
millimeter wavelengths were used to further constrain the parameters of 
the model. The density structure of the core was determined by the SLS 
collapse solution, characterized by a stellar mass and a mass accretion 
rate, obtained from the fit to the observed SED. These two parameters also 
determine the source of heating ($L_* + L_{\rm acc}$) and allow to derive 
the temperature gradient inside the core.

In this way, Osorio et al. (1999) show that the dust continuum emission of 
the HMCs can be explained as arising in massive envelopes collapsing onto 
early spectral type (B) stars with high mass accretion rates 
($10^{-4}$-$10^{-3}~M_\odot$~yr$^{-1}$). In these objects the accretion 
luminosity is the main source of heating. This work shows that massive 
stars up to a mass of 20 $M_{\odot}$ can be formed via accretion since the 
radiation pressure does not halt the collapse.

\section{Flattened  Envelopes}  

\begin{figure}[!t]
%\plotfiddle{osoriofig1.eps}{7.5cm}{0}{40}{40}{-110}{-70}
\plotone{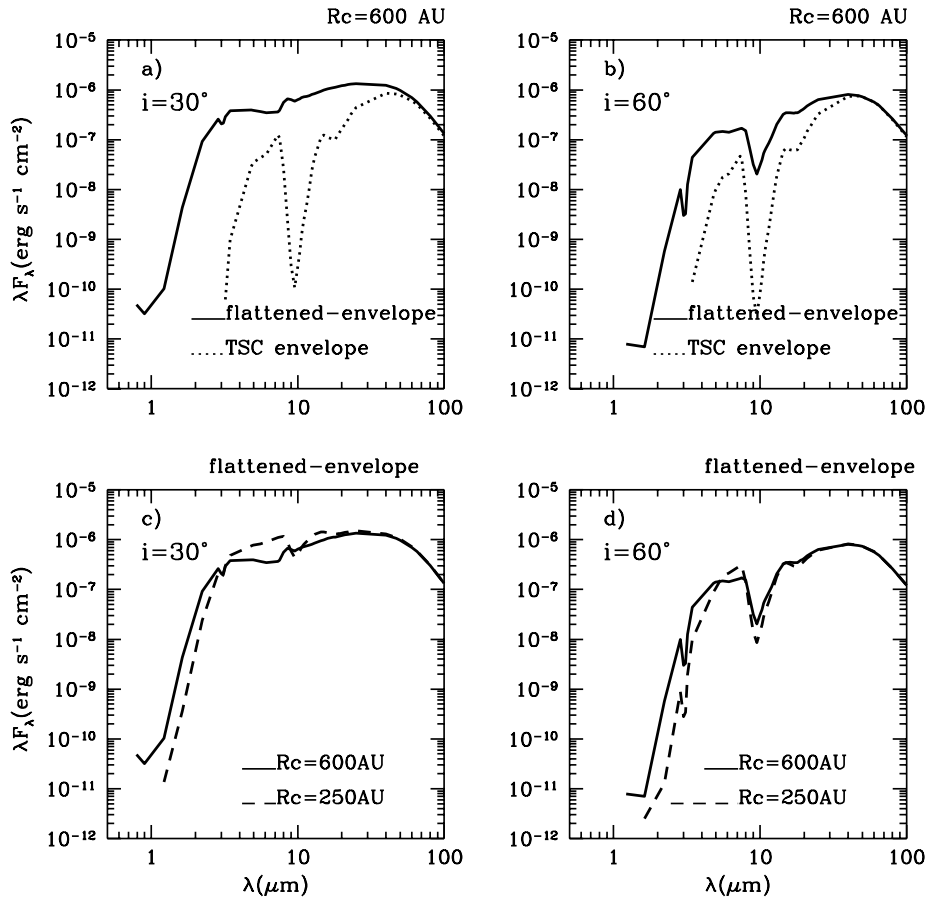}
 \caption{(a) Model SEDs for an intrinsically flattened envelope (solid 
line) and a TSC envelope (dotted line) with the same value of the 
centrifugal radius ($R_c=600$ AU) and inclination angle ($i=30^{\circ}$). 
(b) Same as (a), but for a higher value of the inclination angle 
($i=60^{\circ}$).  (c) Model SEDs for two intrinsically flattened 
envelopes, with the same value of the inclination angle ($i=30^{\circ}$)  
and two different values of the centrifugal radius:  $R_c=250$ AU (dashed 
line) and $R_c=600$ AU (solid line). (d) Same as (c), but for 
$i=60^{\circ}$. The stellar luminosity ($L_*=25000~L_{\odot}$), the mass 
of the envelope (M$_{\rm env}=9~M_{\odot}$), and the assumed distance (1 
kpc) are the same in all models (adapted from De Buizer et al. 2005).
 \label{fig:sedgrid}}
\end{figure}

De Buizer et al. (2005) imaged the near- and mid-infrared emission 
associated with high-mass protostellar objects using the Gemini telescope 
with high angular resolution ($<1''$).  Unfortunately, the SED of these 
sources was not well sampled in the millimeter wavelength range, which is 
very sensitive to the density and luminosity of the sources, and high 
angular resolution millimeter data was available for only one of the 
sources. Nevertheless, since the mid-infrared range is very sensitive to 
the geometry of the source (see Fig.~{\ref{fig:sedgrid}}) it was possible 
to study departures from spherical symmetry using more realistic models. 
Elongated envelopes with flattening due to rotation at small radii and 
with intrinsic flattening at large radii were adopted. This kind of 
envelope is similar to the Terebey et al. (1984) solution (TSC envelope) 
in the inner region but it has been modified at larger radii (Hartmann et 
al. 2006) to simulate more realistically the shape of the star-forming 
core. In that work, parameters such as the inclination of the system, the 
centrifugal radius (where the rotation becomes important), the luminosity, 
and the mass accretion rate are derived.

\begin{figure}[!t]
\plotfiddle{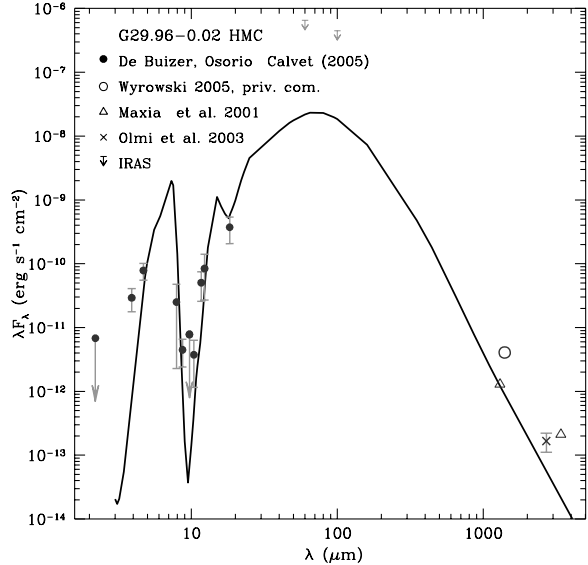}{7.5cm}{0}{40}{40}{-110}{-70}
%\plotone{osoriofig2.eps}
 \caption{Model SED for the source G29.96-0.02 HMC.  The different symbols 
represent the observed values of the flux density. Solid line represents 
the best fit model, obtained with $i=12^{\circ}$, $R_c=570$ AU, 
$L_*=1.8\times 10^{4}~L_{\odot}$, and $\rho_{\rm 1\,AU}=3 \times 10^{-11}$ 
g cm$^{-3}$. The adopted distance is 8.4 kpc (adapted from De Buizer et 
al. 2005).
 \label{fig:g29}}
\end{figure}

As an example, in Figure {\ref{fig:g29}} we show the observed and model 
SEDs for the prototypical HMC near the UCHII region G29.96-0.02 (G29 HMC). 
This is one of few hot cores with both mid-infrared and millimeter 
emission data. The SED of G29 HMC can be explained by an early type (B) 
star with a very high mass accretion rate ($\sim$ $10^{-2}~M_{\odot}$ 
yr$^{-1}$), a centrifugal radius of the order of 600 AU and an inclination 
angle near to the pole-on position. It is worth noting that the 
centrifugal radius (i.e., the radius where the formation of disks is 
expected to occur) found for G29 HMC is in good agreement with the radius 
obtained from high-angular resolution observations of the disk in Ceph A 
HW2, one of best examples of a disk around a high-mass protostar (Patel et 
al. 2005, Torrelles et al. 2007).

A grid of SEDs for high-mass protostars, assuming rotationally flattened 
envelopes with disks and outflows, has been presented by Robitaille et al. 
(2007). However, the observational dataset of HMCs is still scarce and 
their SEDs from near-infrared to millimeter wavelengths are not well 
covered, making difficult to constrain for these objects the large number 
of free parameters of these models.

\section{Modeling of the Molecular Emission: Ammonia Lines}  

The SED alone cannot provide a full description of the physical properties 
of the HMCs. Complementary information can be obtained from the modeling 
of the molecular emission that is sensitive to velocity motions such as 
infall, turbulence, or rotation. Osorio et al. (2007) present a modeling 
procedure aimed to reproduce simultaneously both the SED and the ammonia 
line emission of HMCs.  The only free parameter in this molecular modeling 
is the ammonia gas-phase abundance.

This modeling procedure has been applied to the hot core near the UCHII 
region G31.41+0.31 (G31 HMC). G31 HMC has a quite wide observational set 
of dust emission data, and has been observed in several ammonia inversion 
transitions. In particular, the NH$_3$(4,4) transition has been 
observed with the VLA with subarsecond angular resolution (Cesaroni et al. 
1998). These observations reveal variations of the ammonia emission as a 
function of distance to the center of the core.

\begin{figure}[!tb]
%\epsscale{0.5}
\plotfiddle{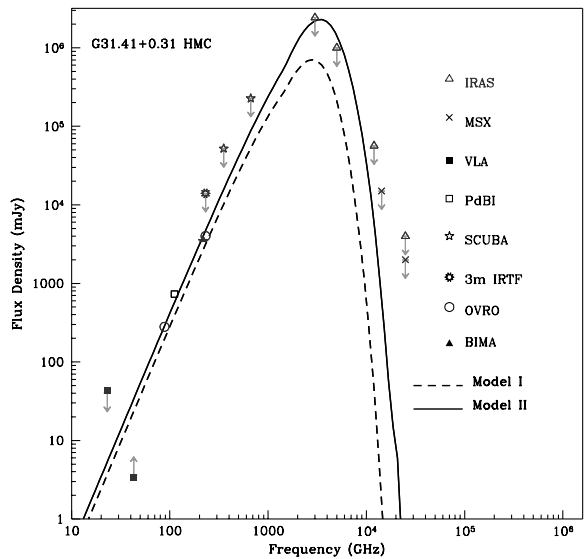}{7.5cm}{0}{40}{40}{-110}{-70}
%\plotone{osoriofig3.eps}
\caption{Observed flux densities of G31 HMC and
predicted SED for Model~I (dashed line) and Model~II (solid line).
(adapted from Osorio et al. 2007).
 \label{fig:G31}}
\end{figure}

\begin{figure}[!bt]
\plotfiddle{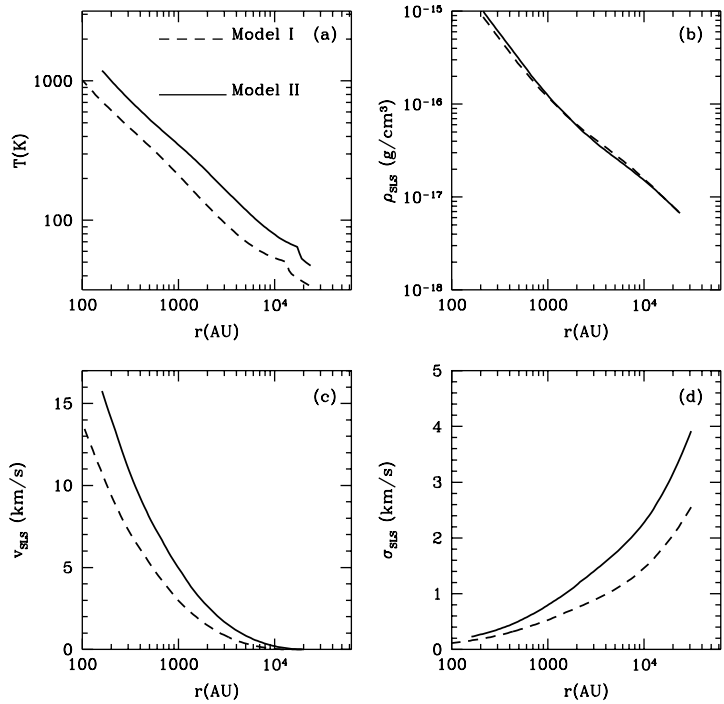}{9.6cm}{0}{50}{50}{-160}{-70}
%\plotone{osoriofig4.eps}
 \caption{Physical structure of the G31 HMC for Model I (dashed line) and
Model II (solid line line).  (a) Dust temperature as a function of radius;
(b) Gas density as a function of radius; (c) Infall velocity as a function
of radius; (d) Turbulent velocity dispersion as a function of radius.
(adapted from Osorio et al. 2007).
 \label{fig:distr}}
\end{figure}

In order to calculate the molecular emission, the density, temperature, 
velocity, and velocity dispersion inside the core are required. Osorio et 
al. (2007) fitted the SED of G31 HMC adopting the physical structure of 
the SLS collapse. Due to the incompleteness of the observational data, two 
models were found to be consistent with the observations (see 
Fig.~{\ref{fig:G31}}). Model I has a stellar mass of $12~M_{\odot}$ and a 
mass accretion rate of $1.6\times 10^{-3}~M_{\odot}$ yr$^{-1}$ and Model 
II has a stellar mass of $25~M_{\odot}$ and a mass accretion rate of $2.7 
\times 10^{-3}~M_{\odot}$ yr$^{-1}$, which results in a more luminous 
envelope ($2.3 \times 10^5~L_{\odot}$) than Model I ($5.0 \times 
10^4~L_{\odot}$). Therefore Model II has a higher temperature than Model I at 
any radius (see Figure {\ref{fig:distr}}).

For a given physical structure of the core the gas-phase ammonia abundance 
is the only free parameter to calculate the ammonia emission. As a first 
approach, a constant gas-phase ammonia abundance inside the core was 
assumed, and a grid of cases was run for a wide range of values of the 
ammonia abundance. However, none of the two models could reproduce the 
VLA ammonia (4.4) spectra obtained by Cesaroni et al. (1998).

As a second approach, Osorio et al. (2007) considered that the total 
abundance of ammonia molecules as a function of radius remains constant 
inside the envelope (no chemical effects) but the ratio of solid to 
gas-phase molecules changes as a function of density and temperature 
inside the core, being described by an equation of thermal balance 
between sublimation and condensation (Sandford \& Allamandola 1993). This 
appears to be a more realistic description since it is expected that 
ammonia molecules are trapped in water ice mantles of dust grains in the 
outer (colder) regions of the core being released to the gas phase in the 
inner (hotter) regions where water molecules are sublimated. Therefore, a 
rapid enhancement of the gas-phase ammonia abundance is expected at the 
radii where temperatures above $\sim100$ K (the sublimation temperature of 
water, for the typical densities of HMCs) are reached. The maximum and 
minimum gas-phase ammonia abundances are, thus, the free parameters to be 
adjusted in this line fitting.

\begin{figure}[!tb]
\plottwo{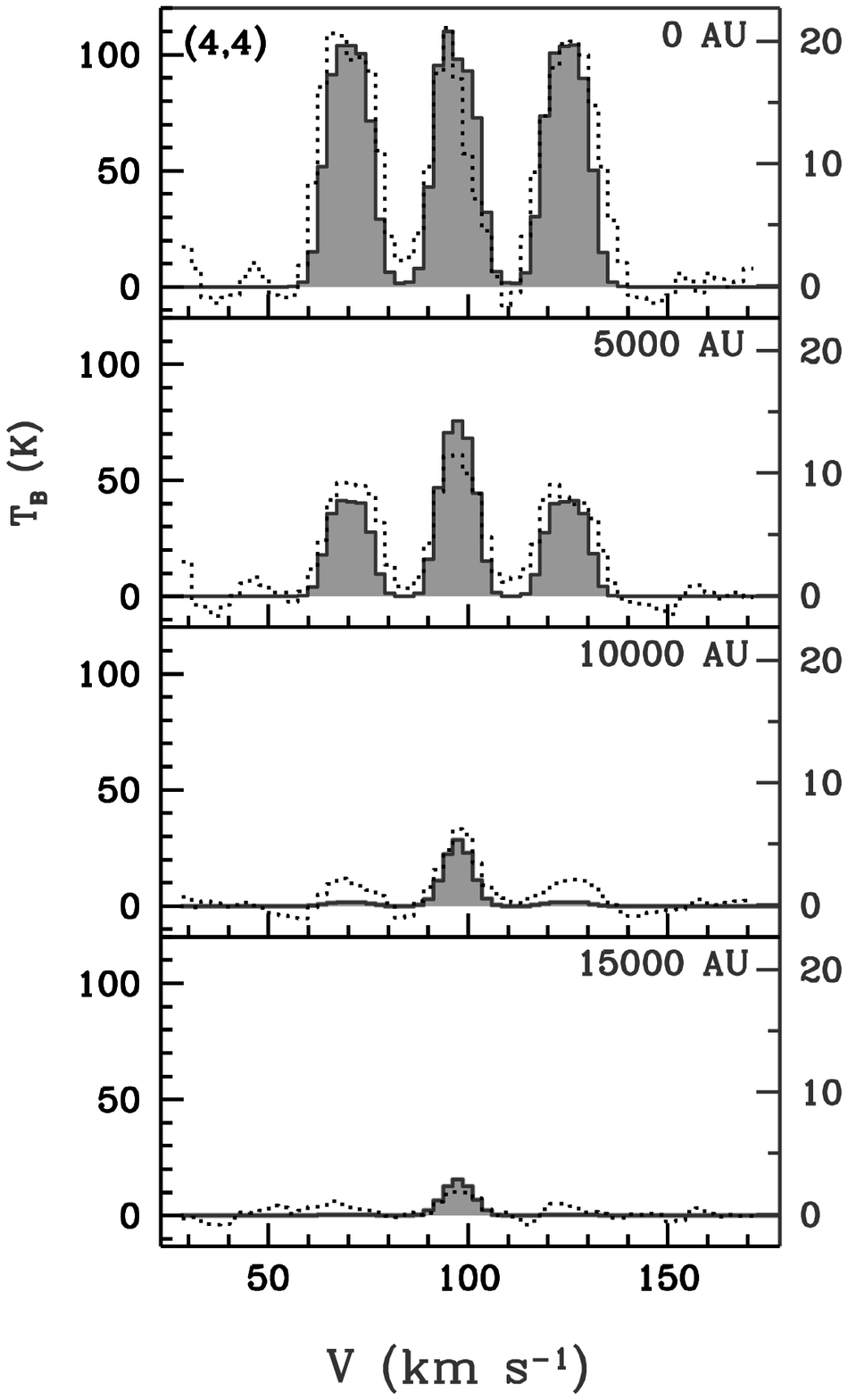}{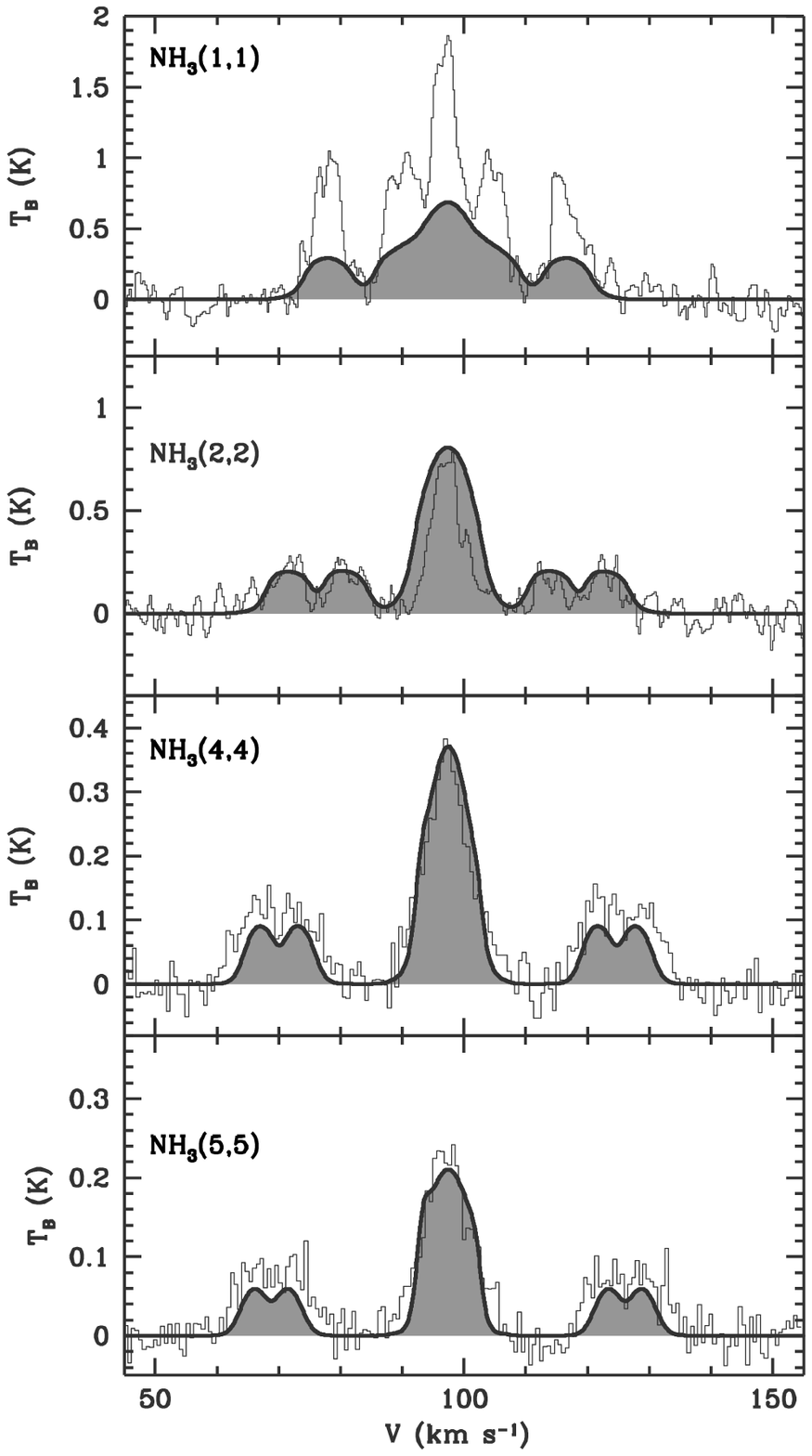}
\caption{
 {\em (Left)} Synthetic spectra of the NH$_3(4,4)$ transition (solid line 
with gray area) for Model II as a function of the projected distance to 
center of the G31 HMC core. The spectra have been obtained assuming a 
variable gas-phase NH$_3$ abundance along the envelope with a minimum 
value of $2\times10^{-8}$ and a maximum value of $3\times10^{-6}$. The 
observed spectra (adapted from Fig. 9c of Cesaroni et al. 1998) are 
plotted in each panel as dotted lines. The angular resolution is 
$0\rlap.''63$. {\em (Right)} Synthetic spectra (solid line with gray area) 
of the NH$_3(1,1)$, NH$_3(2,2)$, NH$_3(4,4)$, and NH$_3(5,5)$ transitions 
towards the center of the HMC, for an angular resolution of $40''$. The 
model parameters and ammonia abundances are the same as in the left 
panels. The spectra observed with the 100 m telescope (Cesaroni et al. 
1992) are also shown (thin line). To facilitate comparison, the observed 
(2,2) and (5,5) spectra have been scaled down 30\% (the uncertainty in 
the absolute calibration of the observed spectra). (adapted from Osorio et 
al. 2007).
 \label{fig:G31h2o}}
\end{figure}

After running a grid of cases it was found that Model I was unable to fit 
the VLA ammonia (4,4) data. However, a reasonably fit was found for Model 
II. This fit reproduces the observed intensity of the main and satellite 
lines as a function of the projected distance to the center, as well as 
the observed line widths (see Fig.~\ref{fig:G31h2o}, left). In this best 
fit, the gas-phase ammonia abundance has a minimum value of $2 \times 
10^{-8}$, typical of the average values reported in the literature for 
cold cores, and a maximum value of $3 \times 10^{-6}$, typical of the 
average values reported for massive cores.

The set of physical parameters derived from the fit to the SED and the 
ammonia abundances derived from the fitting of the VLA NH$_3(4,4)$ spectra 
can also reproduce satisfactorily other ammonia transitions. 
Figure~\ref{fig:G31h2o} (right) shows the spectra of the NH$_3$(1,1), 
NH$_3$(2,2), NH$_3(4,4)$, and NH$_3(5,5)$ transitions observed with the 
100 m telescope (Cesaroni et al. 1992). To facilitate comparison, the 
observed (2,2) and (5,5) spectra have been scaled down 30\% (the 
uncertainty in the absolute calibration of the observed spectra). As can 
be seen in the figure, the model reproduces quite well the observed 
spectra, except for the NH$_3$(1,1) transition, were there is likely a 
contribution of cold molecular gas from outside the core, which is not 
considered in our modeling.

\section{Conclusions}  

\begin{itemize}

\item A spherically symmetric model of the collapse of a SLS can explain
the observed SED and the intensity spatial profiles of the continuum dust 
emission of HMCs, implying that these objects are dominated by accretion.

\item In order to fit the data a young, early type central star with a 
high mass accretion rate is required. These results strengthen the 
hypothesis that HMCs are one of the earliest observable phases of massive 
star formation.

\item Inclusion of rotation and the natural elongation of the cloud allows to 
fit the high angular resolution mid-IR data providing a determination of 
additional physical parameters such as the inclination angle or the 
centrifugal radius. Values of a few hundred AUs are found for this radius, 
similar to those obtained in high angular resolution observations of disks 
around massive protostars.

\item The ammonia emission and its variation across the core can be 
reproduced in great detail provided the variation of the gas-phase ammonia 
abundance due to sublimation of the ammonia molecules from ice grain 
mantles because of the temperature gradient inside the core is taken into 
account.

\item This kind of modeling would be required to explain the details 
of the observational data that are expected to come from the new generation 
of high angular facilities (EVLA, ALMA,...).

\end{itemize}

\acknowledgements %%% Text of acknowledgements runs on after this command.
 M. O. acknowledges support from grant AYA 2005-08523-C03 of the Spanish 
MEC (cofunded with FEDER funds) and from Junta de Andaluc\'\i a.

%%% THE BIBLIOGRAPHY
%%%
%%% CONSULT SECTION 3 OF "INSTRUCTIONS FOR AUTHORS" FOR HOW TO USE NATBIB.
%%% AUTHORS ARE ENCOURAGED TO USE EITHER THE "THEBIBLIOGRAPY" ENVIRONMENT
%%% BY UNCOMMENTING (DELETING THE "%" SYMBOL) THE COMMANDS BELOW, OR BY
%%% USING THE BIBTEX ENVIRONMENT. TO FIND OUT WHICH IS APPLICABLE TO YOUR
%%% CONTRIBUTION, CONSULT THE VOLUME EDITORS FOR YOUR PROCEEDINGS.
%%%


\begin{thebibliography}{}

\bibitem[Cesaroni et al.(1998)]{1998A&A...331..709C} Cesaroni, R., Hofner, 
P., Walmsley, C.~M., \& Churchwell, E.\ 1998, \aap, 331, 709 

\bibitem[Cesaroni, Walmsley, \& Churchwell (1992)]{CWCH92}
 Cesaroni, R., Walmsley, C. M., \& Churchwell, E. 1992, A\&A, 256, 618

\bibitem[De Buizer et al.(2005)]{2005ApJ...635..452D} De Buizer, J.~M., 
Osorio, M., \& Calvet, N.\ 2005, \apj, 635, 452 

\bibitem[Hartmann et al.(1996)]{1996ApJ...464..387H} Hartmann, L., Calvet, 
N., \& Boss, A.\ 1996, \apj, 464, 387 

\bibitem[McLaughlin \& Pudritz(1997)]{1997ApJ...476..750M} McLaughlin, 
D.~E., \& Pudritz, R.~E.\ 1997, \apj, 476, 750 

\bibitem[Osorio et al.(2007)]{2007Oso} Osorio, M., Anglada, G., Lizano, 
S., \& D'Alessio, P.\ 2007, ApJ, submitted

\bibitem[Osorio et al.(1999)]{1999ApJ...525..808O} Osorio, M., Lizano, S., 
\& D'Alessio, P.\ 1999, \apj, 525, 808 

\bibitem[Patel et al.(2005)]{2005Natur.437..109P} Patel, N.~A., et al.\ 
2005, \nat, 437, 109 

\bibitem[Sandford \& Allamandola(1993)]{1993ApJ...417..815S} Sandford, 
S.~A., \& Allamandola, L.~J.\ 1993, \apj, 417, 815 

\bibitem[Robitaille et al.(2007)]{2007ApJS..169..328R} Robitaille, T.~P., 
Whitney, B.~A., Indebetouw, R., \& Wood, K.\ 2007, \apjs, 169, 328 

\bibitem[Terebey et al.(1984)]{1984ApJ...286..529T} Terebey, S., Shu, 
F.~H., \& Cassen, P.\ 1984, \apj, 286, 529 

\bibitem[Torrelles et al.(2007)]{2007ApJ...666L..37T} Torrelles, J.~M., 
Patel, N.~A., Curiel, S., Ho, P.~T.~P., Garay, G., \& Rodr{\'{\i}}guez, 
L.~F.\ 2007, \apjl, 666, L37 

\end{thebibliography}
\end{document}